\def\sqr#1#2{{\vcenter{\vbox{\hrule height.#2pt\hbox{\vrule
width.#2pt height#1pt \kern#1pt\vrule width.#2pt}\hrule height.#2pt}}}}
\def\square{\mathchoice\sqr54\sqr54\sqr{2.1}3\sqr{1.5}3}

\magnification\magstep1
\font\cst=cmr10 scaled \magstep3
\font\csc=cmr10 scaled \magstep2
\vglue 2cm
\centerline{\cst  Linearized gravity on branes~:}
\vskip 0.5cm
 \centerline{\cst from Newton's law to cosmological perturbations}

\vskip 1.5 true cm
\centerline{Nathalie Deruelle}
\vskip 1 true cm
\centerline{\it  Institut d'Astrophysique de Paris,}
\centerline{\it GReCO,  FRE 2435  du CNRS,}
\centerline{\it 98 bis boulevard  Arago, 75014, Paris, France}

\medskip
\vskip 0.5cm
\centerline{November 2002}

\vskip 0.5cm
\centerline{Contribution to the Proceedings of the Menorca ERE}

\vskip 2cm
\noindent
{\bf Abstract}
\bigskip
 We review here how Newton's
law can be approximately recovered in the simple, ``paradigmatic", case of a flat $Z_2$-symmetric brane in a 5-D anti-de Sitter bulk. We then comment on the difficulties
encountered so far in extending this analysis to cosmological perturbations on a Robertson-Walker brane.

\vfill\eject

\noindent
{\csc I. Introduction}
\medskip
Since the seminal papers by Randall and Sundrum [1] there has been an increasing interest for gravity theories within spacetimes with one large extra dimension and the idea
that our universe may be a four dimensional singular hypersurface, or ``brane", in a five dimensional spacetime, or ``bulk". The Randall-Sundrum scenario [1], where our  universe
is a four dimensional quasi-Minkowskian edge of a double-sided perturbed anti-de Sitter spacetime, was the first explicit model where the linearized Einstein equations were
found to hold on the brane, apart from small $1/r^2$ corrections to Newton's potential. This claim was thoroughly analyzed and the corrections to Newton's law exactly
calculated  [2]. Cosmological models were then built, where the brane is taken to be a Robertson-Walker spacetime embedded in an anti-de Sitter bulk [3]. The perturbations of
these models, in the view of calculating the microwave background anisotropies, are currently being studied and compared to the perturbations of standard, four dimensional,
Friedmann universes [4]. 

In this contribution we review the approach advocated in [5].

\bigskip
\noindent
{\csc II. The bulk gravitons}
\medskip

In the conformally Minkowskian coordinates $X^A=\{x^\mu=[T, \vec r=(x^1,x^2,x^3)], X^4=w\}$ the line element of a five dimensional perturbed anti-de Sitter spacetime  reads
$$ds^2|_5= {\cal G}_{AB}\,dX^AdX^B\quad\hbox{with}\quad {\cal G}_{AB}=\left({{\cal L}\over
w}\right)^2\,(\eta_{AB}+\gamma_{AB})\eqno(2.1)$$ 
 where $\cal L$ is a (positive) constant and $\gamma_{AB}$ fifteen fonctions of the five coordinates
$X^A$. ${\cal V}_5$ is defined as the part of this perturbed AdS$_5$ spacetime bounded by a 4D timelike hypersurface $\Sigma$ and $w=+\infty$~; $\Sigma$ is such
that the coordinates
$X^A$ cover the whole of ${\cal V}_5$.

 ${\cal V}_5$ is taken to be an Einstein space, that is a solution of  the Einstein equations ${\cal R}_{AB}=-{4\over{\cal L}^2}{\cal
G}_{AB}$ where ${\cal R}_{AB}$ is the Ricci tensor of the metric ${\cal G}_{AB}$. These equations are, at linear order in $\gamma_{AB}$~:
$$\eqalign{{1\over2}&\left[\partial_{AL}\gamma^L_B+\partial_{BL}\gamma^L_A-\partial_{AB}\gamma-\square_5 \gamma_{AB}-
\eta_{AB}(\partial_{LM}\gamma^{LM}-\square_5 \gamma)\right]\cr
-{6\over w^2}&\eta_{AB}\gamma_{ww}
-{3\over2w}\left[\partial_A\gamma_{wB}+\partial_B\gamma_{wA}-\partial_w\gamma_{AB}+
\eta_{AB}(\partial_w\gamma-2\partial_L\gamma^L_w)\right]=0\cr}\eqno(2.2)$$
where all indices are raised with $\eta^{AB}$, where $\gamma=\gamma^L_L$ and where $\square_5=\partial_L\partial^L$. 

 In the gauge/coordinate transformation $X^A\to X^A=\tilde X^A+\epsilon^A(X^B)$, the metric coefficients transform at first order
as $\gamma_{AB}\to\tilde\gamma_{AB}=\gamma_{AB}+\partial_A\epsilon_B+\partial_B\epsilon_A-2\eta_{AB}{\epsilon^4\over w}$ and it is easy to see that one can choose
five functions $\epsilon^A$ such that (with tildes dropped)
$$\gamma_{Aw}=0\,.\eqno(2.3)$$
These conditions reduce the fifteen metric coefficients  to  the ten  $\gamma_{\mu\nu}$ but do not fix the gauge completely, as they do not uniquely determine the five
functions $\epsilon^A$. Indeed, if we perform the further coordinate transformation $X^A\to X^A=\bar X^A+\epsilon^A$ with $\epsilon^4=wd$ and
$\epsilon_\mu=-{1\over2}w^2\partial_\mu d+c_\mu$, where $d(x^\nu)$ and $c_\mu(x^\nu)$ are five arbitrary fonctions of the four coordinates $x^\nu$, then the new metric
coefficients  satisfy $\bar\gamma_{Aw}=0$ as well, and
$$\bar\gamma_{\mu\nu}=\gamma_{\mu\nu}-w^2\partial_{\mu\nu} d+\partial_\mu c_\nu+\partial_\nu c_\mu-2\eta_{\mu\nu}d\,.\eqno(2.4)$$

In the (hence) {\it class} of coordinate systems (2.3-4) the Einstein linearised equations (2.2) reduce to (with tildes dropped)
$$\eqalign{\partial_{\rho\sigma}\gamma^{\rho\sigma}-\square_4\gamma+{3\over
w}\partial_w\gamma&=0\quad,\quad\partial_w(\partial_\rho\gamma^\rho_\mu-\partial_\mu \gamma)=0\quad,\quad\partial_{ww}\gamma-{1\over w}\partial_w\gamma=0\cr
\square_4\gamma_{\mu\nu}+\partial_{ww}\gamma_{\mu\nu}-{3\over w}&\partial_w\gamma_{\mu\nu}=
\partial_{\mu\rho}\gamma^\rho_\nu+\partial_{\nu\rho}\gamma^\rho_\mu-\partial_{\mu\nu}\gamma+
{\eta_{\mu\nu}\over w}\partial_w\gamma\cr}\eqno(2.5)$$
where now all indices are raised with $\eta^{\mu\nu}$, $\gamma\equiv\gamma^\mu_\mu$  and $\square_4\equiv\partial\mu\partial^\mu$.

The first three, constraint, equations are easily solved~: 
$$\gamma=-{1\over6}w^2\partial_\mu D^\mu+C\qquad,\qquad
\partial_\rho\gamma^\rho_\mu=-{1\over6}w^2\partial_{\mu\rho}D^\rho+\partial_\mu C+D_\mu\eqno(2.6)$$ where $C(x^\nu)$ and $D_\mu(x^\nu)$ are five arbitrary functions of
the four coordinates $x^\nu$.

 Now these five functions do not describe any perturbation of the geometry and can be chosen at will. In particular they can be set to zero. Indeed if we perform a coordinate
change which transforms the metric coefficients acccording to (2.4), and choose $d$ and $c_\mu$ such that $\square_4 d=-{1\over6}\partial_\mu D^\mu$ and
$\partial_{\mu\rho}c^\rho+\square_4c_\mu-2\partial_\mu d=-\partial_\mu C-D_\mu$, then in the new, barred, coordinate system~:
$\bar\gamma=\partial_\rho\bar\gamma^\rho_\mu=0$. Hence  the particular,  tranverse traceless system, such that (with bars dropped)
$$\gamma_{ww}=0\qquad,\qquad\gamma_{w\mu}=0\qquad,\qquad\gamma^\rho_\rho=0\qquad,\qquad\partial_\rho\gamma^\rho_\mu=0\eqno(2.7)$$
solves the constraint equations and reduces the ten metric coefficients $\gamma_{\mu\nu}(X^A)$ to five, which represent the five degrees of freedom of AdS$_5$ gravitational
waves/gravitons.

As for the fourth, evolution, equation (2.5) it reduces in the gauge (2.7) to
$$\square_4\gamma_{\mu\nu}+\partial_{ww}\gamma_{\mu\nu}-{3\over w}\partial_w\gamma_{\mu\nu}=0\,.\eqno(2.8)$$

\medskip
\noindent
$\bullet$ {\it Remark on gaussian normal gauges}.  Another gauge is frequently used~: a gaussian normal gauge which belongs to the class (2.3), so that its metric
coefficients are related to the transverse traceless ones by (2.4), but which is adapted to the boundary $\Sigma$ of ${\cal V}_5$, whose equation, in the coordinate system (2.7)
is $w={\cal L}+\zeta(x^\mu)$.   This gaussian normal coordinate system $X^{(G)A}$ is defined at linear order in
$\zeta$ by
$X^{(G)\mu}= x^\mu+{w^2\over2{\cal L}}\partial_\mu\zeta-{1\over2}{\cal L}\partial_\mu\zeta$,
and $w^G=w(1-{\zeta\over{\cal L}})$. In that gauge the metric coefficients are related to the transverse traceless ones (2.7) (2.8) by
$$\gamma^G_{\mu\nu}=\gamma_{\mu\nu}-\left({w^2\over{\cal L}^2}-1\right)\partial_{\mu\nu}\zeta-2\eta_{\mu\nu}{\zeta\over{\cal L}}\,.\eqno(2.9)$$
They can be obtained alternatively by  choosing $D_\mu={6\over{\cal L}^2}\partial_\mu\zeta$ and $C=\square_4\zeta-{8\zeta\over{\cal L}}$, so that their
evolution equation (2.5) is
$$\square_4\gamma^G_{\mu\nu}+\partial_{ww}\gamma^G_{\mu\nu}-{3\over
w}\partial_w\gamma^G_{\mu\nu}=-\left({w^2\over{\cal L}^2}-1\right)\partial_{\mu\nu}\square_4\zeta+{4\over{\cal
L}^2}\partial_{\mu\nu}\zeta-2{\eta_{\mu\nu}\over{\cal L}^2}\square_4\zeta\eqno(2.10)$$
whose general solution is (2.9). An important feature of this gaussian normal gauge is that the metric coefficients diverge for large $w$. At linear order in $\zeta$ this gauge
artefact is well under control, but becomes a nuisance when one treats more elaborate models, e.g. cosmological branes or black holes on branes, as the boundary condition on
the metric coefficients is no longer that they must converge when $w\to\infty$. We shall therefore stick to the gauge (2.7).

\medskip
\noindent
$\bullet$ {\it  AdS$_5$ {\sl vs} bulk gravitons}. The general solution of (2.8) depends on the boundary conditions and, as usual, we impose that it converges
(more precisely is
$L_2$) on its domain of definition. If the spacetime we consider spans the whole $w$ axis then
$$\gamma_{\mu\nu}(X^A)={\cal R}e\int\! {d^3k\over(2\pi)^{3\over2}}\,{dm\over(2\pi)^{1\over2}}\,{\rm e}^{{\rm i}k_\rho X^\rho}e_{\mu\nu}\,w^2Z_2(mw)\quad
\hbox{with}\quad \left\{\eqalign {& k_0=-\sqrt{k^2+m^2}\cr
& k^\rho e_{\rho\mu}=0\cr
&\eta^{\rho\sigma}e_{\rho\sigma}=0\cr}\right.
\eqno(2.11)$$
where the five polarisations $e_{\mu\nu}$ are a priori arbitrary functions of $k^i$ and $m$ and where $Z_2(mw)=H^{(1)}_2(mw)+a_mH^{(2)}_2(mw)$ is an a priori arbitrary linear
combination of second order Hankel functions of first and second kinds. The coefficient $a_m$ is determined by the model at hand~ and  one usually eliminates the
mode coming from $+\infty$, that is sets $a_m=0\,$. The ``zero modes" are the particular, bounded,  solutions of (2.8) which do not depend on $w$
$$\gamma^{(0)}_{\mu\nu}(x^\mu)={\cal R}e\int\! {d^3k\over(2\pi)^{3\over2}}\,{\rm e}^{{\rm i}(k_i x^i-kT)}e_{\mu\nu}(k^i)\qquad
\hbox{with}\qquad \left\{\eqalign {& ke_{0\mu}+k^ie_{i\mu}=0\cr
&\eta^{\rho\sigma}e_{\rho\sigma}=0\,.\cr}\right.\eqno(2.12)$$
 Now, if the spacetime we consider is ${\cal V}_5$ (which is delineated by the hypersurface $\Sigma$ and $w\to+\infty$) then $w$
is bounded by $\Sigma$ and does not go to
$-\infty$. Equation (2.8) then possesses extra $L_2$ modes which converge exponentially as $w\to+\infty$. The general solution of (2.8) in ${\cal V}_5$ is therefore the sum of
(2.11) and 
$$\gamma_{\mu\nu}(X^A)={\cal R}e\!\int\! {d^3k\over(2\pi)^{3\over2}}\int_{-k^2}^0\!{dA\over(2\pi)^{1\over2}}\,{\rm e}^{{\rm i}k_\rho
X^\rho}\!e_{\mu\nu}\,w^2H^{(1)}_2({\rm i}\sqrt{|A|}\,w)\quad
\hbox{with}\ \left\{\eqalign {& k_0=-\sqrt{k^2+A}\cr
& k^\rho e_{\rho\mu}=0\cr
&\eta^{\rho\sigma}e_{\rho\sigma}=0\,.\cr}\right.
\eqno(2.13)$$
The static modes are the particular solutions which do not depend on time~:
$$\gamma^{(s)}_{\mu\nu}(x^i,w)={\cal R}e\int\! {d^3k\over(2\pi)^{3\over2}}{\rm e}^{{\rm i}k_i x^i}e_{\mu\nu}(k^i)\,w^2H^{(1)}_2({\rm i}kw)\,.\eqno(2.14)$$

\bigskip
\noindent
{\csc III.  The equations for gravity on a quasi-minkowskian brane}
\medskip
We consider in AdS$_5$ the hypersurface $\Sigma$ defined, in the coordinate system (2.7), by
$$w={\cal L}+\zeta(x^\mu)\eqno(3.1)$$
where the function $\zeta(x^\mu)$ is a priori arbitrary and describes the so-called ``brane-bending" effect.
The induced metric on $\Sigma$ is 
$$ds^2=(\eta_{\mu\nu}+h_{\mu\nu})dx^\mu dx^\nu\quad\hbox{with}\quad h_{\mu\nu}=\gamma_{\mu\nu}|_{\Sigma} -
2{\zeta\over{\cal L}}\eta_{\mu\nu}\eqno(3.2)$$
where $\gamma_{\mu\nu}(x^\mu,w)$ is a solution of (2.7) (2.8) and where the index $\Sigma$ means that the quantity is evaluated at $w={\cal L}$. (Alternatively this induced
metric can be obtained using the  gaussian normal system introduced in (2.9), in which the equation of $\Sigma$ is $w^G={\cal L}$ and
$h_{\mu\nu}=\gamma^G_{\mu\nu}|_\Sigma$.)

The Randall-Sundrum brane scenario is obtained by cutting AdS$_5$ along $\Sigma$, by making a copy of the
$w\geq{\cal L}+\zeta$ side and pasting it along  $\Sigma$. Imposing that the linearised Einstein equations be valid everywhere in this new ``bulk" manifold, including its edge, or
brane, yields the Israel junction conditions which give the
stress-energy tensor of the matter in the brane $\Sigma$ as
$\kappa T^\mu_\nu=-{6\over{\cal L}}\delta^\mu_\nu+\kappa  S^\mu_\nu$
with $\kappa$ a coupling constant and
$${\kappa\over2}\left( S_{\mu\nu}-{1\over3}\eta_{\mu\nu} S\right)=\partial_{\mu\nu}\zeta-{1\over2}
(\partial_w\gamma_{\mu\nu})|_\Sigma\,.\eqno(3.3)$$
Equations (3.2) and (3.3)  together with (2.7) and (2.8) completely describe gravity in the brane. They have two useful consequences
$$\partial_\rho S^\rho_\nu=0\qquad,\qquad \square_4\zeta=-{\kappa\over6}S\,.\eqno(3.4)$$

Let us compare them  to the standard 4D Einstein equations. In order to do so, we first perform a coordinate transformation  $x^\mu\to x^\mu=
x^{*\mu}+\epsilon^\mu$. Then the new metric coefficients
$$h^*_{\mu\nu}=h_{\mu\nu}|_\Sigma-{2\over{\cal L}}\zeta\eta_{\mu\nu}+
\partial_\mu\epsilon_\nu+\partial_\nu\epsilon_\mu\quad\hbox{with}\quad\square_4\epsilon_\mu=-{2\over{\cal L}}\partial_\mu\zeta
\eqno(3.5)$$
 satisfy the harmonicity condition
$$\partial_\mu \left(h^{*\mu}_\nu-{1\over2}\delta^\mu_\nu h^*\right)=0\,.\eqno(3.6)$$
Taking the d'Alembertian of (3.5) and using (3.3) (2.7) and (2.8) we get the following consequence of the brane gravity equations 
$$\square_4h^*_{\mu\nu}=-16\pi G\left(S_{\mu\nu}-{1\over2}\eta_{\mu\nu}S\right)
-(\partial_{ww}\gamma_{\mu\nu})|_\Sigma+{1\over{\cal L}}(\partial_w\gamma{\mu\nu})|_\Sigma\eqno(3.7)$$
where we have identified ${\kappa\over{\cal L}}\equiv 8\pi G$, $G$ being Newton's constant. 

Now, the standard linearized Einstein equations on a
4D Minkowski background read, in harmonic coordinates
$$\square_4h^*_{\mu\nu}=-16\pi G\left(S_{\mu\nu}-{1\over2}\eta_{\mu\nu}S\right)\,.\eqno(3.8)$$
They hold for {\it any} type of matter (compatible with the harmonicity condition, or, equivalently with the Bianchi identity $\partial_\rho
S^\rho_\nu=0$). By contrast, the linearized equations for gravity on a brane are (3.7), {\it provided} the source $S_{\mu\nu}$ satisfies the junction condition (3.3). This
proviso may restrict the type of matter which is allowed on the brane~: if, for example, only zero modes are allowed in the bulk, then (3.7) reduces to the Einstein linearized
equations, but, since the last term in (3.3) is absent, the derivatives $\partial_\lambda({\cal S}_{\mu\nu}-{1\over3}\eta_{\mu\nu} {\cal S})$ must be
symmetric in $\lambda$ and $\mu$, a property which is not satisfied by standard matter. 

In order to dwell on that point, let us decompose the stress-energy tensor of matter in Fourier space into the traditional form~:
$S_{\mu\nu}=\int\!{d^3k\over(2\pi)^{3\over2}}e^{{\rm i}k_ix^i}\hat S_{\mu\nu}$ with
$$\left\{\eqalign{\hat S_{00}(t,k^i)=\hat\rho\qquad,&\qquad \hat S_{0i}(t,k^i)=-{\rm i}k_i\hat v-\hat v_i\cr
\hat S_{ij}(t,k^i)=\delta_{ij}&\left(\hat P+{k^2\over3}\hat\Pi\right)-k_ik_j\hat\Pi+{\rm
i}k_i\hat\Pi_j+{\rm i}k_j\hat\Pi_i+\hat\Pi_{ij}\cr}\right.\eqno(3.9)$$ where $k_i\hat
v^i=k_i\hat\Pi^i=k_i\hat\Pi^{ij}=\hat\Pi^i_i=0$. We Fourier transform similarly $\zeta$ and $\gamma_{\mu\nu}$. The junction conditions (3.3) then read
$$\left\{\eqalign{\kappa\hat\rho=&-2k^2\hat\zeta-\partial_w\hat\gamma^l_{l|\Sigma}\quad,\quad\kappa\hat
P={4k^2\over3}\hat\zeta+2\ddot{\hat\zeta}+{2\over3}\partial_w\hat\gamma^l_{l|\Sigma}-\partial_w\hat\gamma_{00|\Sigma}\cr
\kappa\hat\Pi=&2\hat\zeta-{1\over2
k^2}\partial_w\hat\gamma^l_{l|\Sigma}+{3\over2k^4}k^lk^m\partial_w\hat\gamma_{lm|\Sigma} \quad,\quad \kappa\hat v=-2\hat{\dot\zeta}-{\rm i}{k^l\over
k^2}\partial_w\hat\gamma_{0l|\Sigma}\cr
\kappa\hat v_i=&-k_i{k^l\over k^2}\partial_w\hat\gamma_{0l|\Sigma}+
\partial_w\hat\gamma_{0i|\Sigma}\quad,\quad\kappa\hat\Pi_i=-{\rm i}k_i{k^lk^m\over k^4}\partial_w\hat\gamma_{lm|\Sigma}+{\rm i}{k^l\over
k^2}\partial_w\hat\gamma_{il|\Sigma}\cr
\kappa\hat\Pi_{ij}=& {1\over2}\left(\delta_{ij}-{k_ik_j\over k^2}\right)\partial_w\hat\gamma^l_{l|\Sigma}-{1\over2}\left(\delta_{ij}+{k_ik_j\over
k^2}\right){k^lk^m\over k^2}\partial_w\hat\gamma_{lm|\Sigma}\cr &+{k_ik^l\over k^2}\partial_w\hat\gamma_{jl|\Sigma}+ {k_jk^l\over
k^2}\partial_w\hat\gamma_{il|\Sigma}-\partial_w\hat\gamma_{ij|\Sigma}\cr}\right.\eqno(3.10)$$
(which can be further simplified using $\hat\gamma_{00}=\hat\gamma^l_l$ and $-{\rm i}\dot{\hat\gamma}_{0\mu}=k_l\hat\gamma^l_\mu$).
Suppose now that only the zero modes are allowed in the bulk. Then, since $\partial_w\hat\gamma_{\mu\nu|\Sigma}=0$ in
that case, the matter on the brane is forced to obey the very contrived equation of state
$$\hat\rho=-k^2\hat\Pi\quad,\quad \hat v=-\dot{\hat\Pi}\quad,\quad
\hat P=\ddot{\hat\Pi}+{2k^2\over3}\hat\Pi\qquad,\qquad\hat v_i=\hat\Pi_i=\hat\Pi_{ij}=0\,.\eqno(3.11)$$
As for the only free function $\hat\Pi$ it is given in terms of the brane bending function $\hat\zeta$ by~:
$\kappa\hat\Pi=2\hat\zeta$. 

Now, of course, the junction conditions are better seen as boundary conditions on the allowed modes in the bulk. Indeed, if matter on the brane is known, then (3.3) or (3.10) can
be inverted to give $\hat\zeta$ and $\partial_w\hat\gamma_{\mu\nu|\Sigma}$ in terms of $\hat S_{\mu\nu}$. Now, from (2.11) and (2.13) we have
$$\eqalign{\partial_w\hat\gamma_{\mu\nu|\Sigma}=&{\cal R}e\int\!{dm\over\sqrt{2\pi}}\,{\rm e}^{-{\rm i}\sqrt{k^2+m^2}T}\,e_{\mu\nu}\,m{\cal L}^2\,Z_1(m{\cal
L})\cr &+{\cal R}e\int_{-k^2}^0\!{dA\over\sqrt{2\pi}}\,{\rm e}^{-{\rm i}\sqrt{k^2+A}T}\,e_{\mu\nu}\,{\rm i}\sqrt{|A|}{\cal L}^2\,H_1^{(1)}(i\sqrt{|A|{\cal L}})\cr}\eqno(3.12)$$
which, in principle, gives by inversion the polarisations $e_{\mu\nu}$ in terms of the brane matter source, and hence the allowed bulk gravitons. Then the induced metric on the
brane (3.2) is known in terms of the matter variables and can be compared with the usual 4D Einstein result. This programme however has only been completed in the
particular case of a point static source.
\bigskip
\noindent
{\csc IV. The $1/r^2$ correction to Newton's law}
\medskip

Let us concentrate on a static, point-like source
$$S_{00}=M\delta(\vec r)\qquad,\qquad S_{0i}=S_{ij}=0\,.\eqno(4.1)$$

The junction conditions (3.3) or (3.10) then give, using $\hat\delta={1\over(2\pi)^{3/2}}$, $\hat{1\over r}=\sqrt{2\over\pi}{1\over k^2}$ and
$\hat{\partial_{ij}{1\over r}}=-\sqrt{2\over\pi}{k_ik_j\over k^2}$~:
$$\left\{\eqalign{\zeta&=-{\kappa M\over24\pi}{1\over r}\quad,\quad\partial_w\gamma^{(s)}_{00}|_\Sigma=-{2\kappa M\over 3}\delta (\vec r)\cr
\partial_w\gamma^{(s)}_{0i}|_\Sigma&=0\qquad,\qquad
\partial_w\gamma^{(s)}_{ij}|_\Sigma=-{\kappa M\over 3}\delta (\vec r)\delta_{ij}-{\kappa M\over12\pi}\partial_{ij}{1\over r}\,.\cr}\right.\eqno(4.2)$$

Now, the static bulk modes are given by (2.14)~: $\hat\gamma^{(s)}_{\mu\nu}(k^i,w)=e_{\mu\nu}(k^i)\,w^2H_2^{(1)}({\rm i}kw)$ and their $w$-derivatives on $\Sigma$ by
(3.12)~: $\partial_w\hat\gamma^{(s)}_{\mu\nu}(k^i)=e_{\mu\nu}(k^i)\,{\rm i}k{\cal L}^2\,H_1^{(1)}(ik{\cal L})$. Equation (4.2) therefore gives the polarisations in terms of the
brane stress-energy tensor as
$$\left\{\eqalign{e_{00}(\vec k)H_1^{(1)}(ik{\cal L})&=-{2\kappa M\over 3}{1\over(2\pi)^{3\over2}}{1\over ik{\cal L}^2}\quad,\quad e_{0i}(\vec k)H_1^{(1)}(ik{\cal L})=0\cr
e_{ij}(\vec k)H_1^{(1)}(ik{\cal L}))&=-{\kappa M\over 3}
{1\over(2\pi)^{3\over2}}{1\over ik{\cal L}^2}\left(\delta_{ij}-{k_ik_j\over4\pi k^2}\right)\,.\cr}\right.\eqno(4.3)$$
The bulk metric  is then known as~:
$$\gamma_{\mu\nu}(\vec r,w)={\cal R}e\int\!{d^3k\over(2\pi)^{3\over2}}\,e^{i\vec k\,.\vec r}\hat\gamma_{\mu\nu}(\vec k, w)\quad ,\quad\hat\gamma_{\mu\nu}(\vec k,w)=
{\kappa M\over3{\cal L}(2\pi)^{3\over2}}\,w^2\,{K_2(kw)\over k{\cal L}\,K_1(k{\cal L})}\,c_{\mu\nu}\eqno(4.4)$$
with $c_{00}=2$, $c_{0i}=0$ and $c_{ij}=\delta_{ij}-k_ik_j/k^2$ and where $K_\nu(z)$ is the modified Bessel function defined
as $K_\nu(z)=i{\pi\over2}e^{i\nu{\pi\over2}}H_\nu^{(1)}(iz)$. 

Let us now concentrate on the $h_{00}$ component of the metric on the brane (which is the same in the $x^\mu$ coordinates and the harmonic coordinates $x^{*\mu}$). With
$\zeta$ given by (4.2) it reads
$$\hat h_{00}(\vec k)=\hat h^*_{00}(\vec k)=\hat\gamma_{00}|_\Sigma+2{\hat\zeta\over{\cal L}}=
{\kappa M\over k^2{\cal L}(2\pi)^{3\over2}}\left[1-{2k{\cal L}\over3}{K_0(k{\cal L})\over K_1(k{\cal L})}\right]\,.\eqno (4.5)$$
Taking the Fourier transform and integrating over angles we obtain, setting $\alpha=r/{\cal L}$ and recalling that ${\kappa\over{\cal L}}=8\pi G$
$$h_{00}(\vec r)={2GM\over r}\left(1+{4\pi\over3}{\cal K}_\alpha\right)\quad\hbox{with}\quad {\cal K}_\alpha=\lim_{\epsilon\to0}\int_0^{+\infty}\!
du\,\sin(u\alpha)\,{K_0(u)\over K_1(u)}e^{-\epsilon u}\,.\eqno(4.6)$$
 It is a (fairly) straightforward exercise to see that $\lim_{\alpha\to0}{\cal K}_\alpha=\alpha^{-1}={\cal L}/r$ and that $\lim_{\alpha\to\infty}{\cal
K}_\alpha=\pi/2\alpha^2=\pi ({\cal L}/r)^2/2$. At short distances the correction to Newton's law is in ${\cal L}/r$, whereas as distances large compared with the characteristic
scale ${\cal L}$ of the anti-de Sitter bulk the correction is reduced by another ${\cal L}/r$ factor
$$\lim_{r/{\cal L}\to\infty}h_{00}(\vec r)={2GM\over r}\left[1+{2\over3}\left({{\cal L}\over r}\right)^2\right]\,.\eqno(4.7)$$
\bigskip
\noindent
{\csc V. Cosmological branes and their perturbations}
\medskip
Let us now consider in (unperturbed) AdS$_5$ spacetime with line element $ds^2|_5=\left({{\cal L}\over
w}\right)^2\eta_{AB}\,dX^AdX^B$ the hypersurface $\Sigma$ defined by
$$w={{\cal L}\over a(\eta)}\quad,\quad T=\int\!d\eta\,\sqrt{1+{\cal L}^2H^2}\eqno(5.1)$$
with $H\equiv{\dot a\over a}$, a dot denoting differentiation with respect to $t$ such that $dt=a(\eta)d\eta$. $\Sigma$ is then a spatially flat Robertson-Walker 4D spacetime
with scale factor $a(\eta)$ and line element ($x^0=\eta$)
$$ds^2=a^2(\eta)\,\eta_{\mu\nu}dx^\mu dx^\nu\,.\eqno(5.2)$$
One now cuts AdS$_5$ along $\Sigma$,  keeps the part between $\Sigma$ and $w\to+\infty$, copies it and pastes it along $\Sigma$. Imposing the Einstein equations to be
satisfied everywhere in this bulk, including the brane $\Sigma$, yields the Israel junction conditions
$$\kappa\left(S^\mu_\nu-{1\over3}\delta^\mu_\nu S\right)=2K^\mu_\nu\eqno(5.3)$$
where $S_{\mu\nu}$ is the stress-energy tensor of matter on the brane and $K_{\mu\nu}$ the extrinsic curvature of $\Sigma$ in AdS$_5$. For $\Sigma$ defined by (5.1) they
read, setting $S^0_0=-\left({6\over\kappa{\cal L}}+\rho\right)$ and $S^i_j=\delta^i_j\left(-{6\over\kappa{\cal L}}+P\right)$~:
$$\kappa\rho={6\over{\cal L}}\left(\sqrt{1+{\cal L}^2H^2}-1\right)\quad,\quad \dot\rho+3H(\rho+P)=0\,.\eqno(5.4)$$
At late times the first one becomes $8\pi G\rho\to 3H^2$ (with the identification ${\kappa\over{\cal L}}=8\pi G$) and the equations for gravity on the brane become the
standard Friedmann equations. If matter on the brane is imposed to be a scalar field $\Phi(t)$ with potential $V(\Phi)$ and tension ${6\over\kappa{\cal L}}$, the junction
equations (5.3) read
$$\kappa\left({\dot\Phi^2\over2}+V\right)={6\over{\cal L}}\left(\sqrt{1+{\cal L}^2H^2}-1\right)\quad,\quad\ddot\Phi+3H\dot\Phi+{dV\over d\Phi}=0\,.\eqno(5.5)$$

We now allow for gravitons in the bulk and perturb the position of $\Sigma$.  The  equations for $\Sigma$ are then
taken to be
$$w={{\cal L}\over a}+{\zeta\over a}\sqrt{1+{\cal L}^2H^2}\quad,\quad T=\int\!d\eta\,\sqrt{1+{\cal L}^2H^2}-{\zeta\over a}{\cal L}H\,.\eqno(5.6)$$
The line element on $\Sigma$  becomes 
$$ds^2=a^2(\eta)\,(\eta_{\mu\nu}+h_{\mu\nu})\,dx^\mu dx^\nu\eqno(5.7)$$
with
$$\left\{\eqalign{h_{\eta\eta}&=(1+{\cal L}^2H^2)\,\gamma_{00}|_\Sigma
+{2\zeta{\cal L}\over\sqrt{1+{\cal L}^2H^2}}\left({1\over{\cal L}^2}+{\ddot a\over a}\right)\cr
h_\eta^i&=\sqrt{1+{\cal L}^2H^2}\,\gamma_0^i|_\Sigma\quad,\quad h^i_j=\gamma^i_j|_\Sigma-{{2\zeta}\over{\cal L}}\,\sqrt{1+{\cal
L}^2H^2}\,\delta^i_j\,.\cr}\right.\eqno(5.8)$$
where $\gamma_{\mu\nu}|_\Sigma$ is given by (2.11) and (2.13) with $w$ and $T$ given by (5.1). Equations (5.7-8) generalize (3.2) to a Robertson-Walker brane and reduce to it
when $a=1$.

As for the spatial part of the junction conditions (5.3) it becomes, in
the particular case when matter on the brane is imposed to be the perturbed scalar field $\Phi(t)+\chi(\eta,x^i)$ ($\Phi(t)$ solving (5.5))
$$\left\{\eqalign{{\kappa\over2}\delta\left({\cal T}^i_j-{1\over3}\delta^i_j{\cal T}\right)&=
{1\over a^2}\,\partial^i_j\zeta+H\,\delta^i_j\left(H\zeta-\dot\zeta\right)+{1\over2}H^2{\cal L}\,\delta^i_j\,\sqrt{1+{\cal L}^2H^2}\,\gamma_{00}|_\Sigma\,+\cr
{{\cal L}\over2a}&\left[H(\partial_0\gamma^i_j)|_\Sigma-{1\over{\cal L}^2}\sqrt{1+{\cal
L}^2H^2}\,(\partial_w\gamma^i_j)|_\Sigma-H(\partial_j\gamma_0^i+\partial^i\gamma_{0j})|_\Sigma\right]\cr
\hbox{with}\qquad\qquad\qquad&\cr\quad{\kappa\over2}\delta\left({\cal T}^i_j-{1\over3}\delta^i_j{\cal T}\right) & = {\kappa\over
6}\delta^i_j\left[\dot\Phi\dot\chi+\chi{dV\over d\Phi}+{\dot\Phi^2\over2}h_{\eta\eta}\right]\cr}\right.\eqno(5.9)$$
where on the right-hand side spatial indices are raised with $\delta^{ij}$. Equation (5.9) is the generalisation of the spatial part of (3.3) to a cosmological brane and reduces to
it when $a=1$. As for the $(00)$ and $(0i)$ parts of the junction conditions they can be replaced by the Klein-Gordon equation for the scalar field, which  reads 
$$\ddot\chi-{1\over a^2}\Delta\chi+3H\dot\chi+{d^2V\over d\Phi^2}\chi+(\ddot\Phi+3H\dot\Phi)h_{\eta\eta}-{1\over a}\dot\Phi\partial_ih^i_\eta+{\dot\Phi\over2}(\dot
h_{\eta\eta}+\dot h^i_i)=0\,.\eqno(5.10)$$ 

Equations (5.7-10) completely describe the perturbations of a cosmological brane when matter is imposed to reduce to a scalar field.

At that stage, one could put them in a form akin to (3.7) and compare them to the usual equations for cosmological perturbations in 4D Einstein gravity. However, as we have
already seen such equations would be only a consequence of the junction conditions, and not equivalent to them. It is therefore better to stick to (5.7-10).  The junction
conditions (5.9-10) must be seen as boundary conditions giving the polarisations of the bulk gravitons as well as $\zeta$ in terms of $\chi$. The induced metric (5.7-8) is then
in principle also known in terms of $\chi$. Such a programme has however not yet been carried out explicitely (recall that in the much simpler case of a quasi-minkowskian
brane it has been carried out only when matter on the brane is a static point-like source, see Section IV).

One can nevertheless get further insight into them by introducing the spatial tensor
$$F^i_j\equiv {1\over a^2}\,\partial^i_j\zeta+{{\cal L}\over
2a}\left[H(\partial_0\gamma^i_j)|_\Sigma-{1\over{\cal L}^2}\sqrt{1+{\cal
L}^2H^2}\,(\partial_w\gamma^i_j)|_\Sigma-H(\partial_j\gamma_0^i+\partial^i\gamma_{0j})|_\Sigma\right]\eqno(5.11)$$
and casting (5.9) into a traceless and trace part~:
$$\left\{\eqalign{F^i_j&={1\over3}\delta^i_j\, F\cr
F&={\kappa\over
2}\left[\dot\Phi\dot\chi+\chi{dV\over d\Phi}+{\dot\Phi^2\over2}h_{\eta\eta}\right]
-3H(H\zeta-\dot\zeta)-{3\over2}H^2{\cal L}\sqrt{1+{\cal L}^2H^2}\,\gamma_{00}|_\Sigma\,.\cr}\right.\eqno(5.12)$$
 
In doing so, the following, partial but explicit, result can be obtained~: suppose the only modes allowed in the bulk are the zero modes (2.12) and assume (without loss of
generality) that $k_1=k_2=0$, $k_3\equiv k$. The transverse and traceless properties of $e_{\mu\nu}$ then imply that
the five possible polarisations are characterised by $e_{11}$, $e_{12}$, $e_{13}$, $e_{23}$ and $e_{33}$, the other
components being $e_{0i}=-e_{i3}$, $e_{00}=e_{33}$, and $e_{22}=-e_{11}$.

The junction conditions (5.11-12) then tell us first  that $e_{13}$ and $e_{23}$ remain free and correpond to  4D gravitational waves freely propagating in the brane~; they also
tell us that $e_{12}=e_{11}=e_{22}=0$, so that only $e_{33}=-e_{03}=e_{00}\equiv e(k)$ can couple to the brane scalar field~; finally they give
$$\zeta(\eta, z)={\cal R}e\,{{\rm i}Ha{\cal L}\over2}\int\!{dk\over\sqrt{2\pi}}\,{\rm e}^{{\rm i}k(z-T(\eta))}\,{e(k)\over k}\eqno(5.13)$$
with $T(\eta)=\int\!d\eta\,\sqrt{1+{\cal L}^2H^2}$. As for the Klein-Gordon equation (5.10) it  gives
$$\chi(\eta, z)=-{\cal R}e\,{{\rm i}\dot\Phi a\over2}\sqrt{1+{\cal L}^2H^2}\int\!{dk\over\sqrt{2\pi}}\,{\rm e}^{{\rm i}k(z-T(\eta))}\,{e(k)\over k}\,.\eqno(5.14)$$
The metric on the brane is then given by (5.7-8), with $\zeta$ given above and
$$\gamma_{33|\Sigma}(\eta, z)=-\gamma_{03|\Sigma}(\eta, z)=\gamma_{00|\Sigma}(\eta, z)= {\cal R}e\,\int\!{dk\over\sqrt{2\pi}}\,{\rm e}^{{\rm
i}k(z-T(\eta))}\,e(k)\,.\eqno(5.15)$$

Hence the perturbations of this particular cosmological brane are completely known in terms of $e(k)$. Before eventually  comparing them with the standard 4D perturbations of
chaotic inflationary models, one must first decide on the $k$ dependence of $e(k)$. One could try and argue that what matters is matter on the brane and, hence,
that one should impose the field $\chi$  to be in its vacuum state which would amount, in practice, to choosing ${e(k)\over k}\propto {1\over\sqrt{2k}}$. One could on
another hand argue that gravitons in the bulk should be in their vacuum state and impose $e(k)\propto {1\over\sqrt{2k}}$. The right answer to this
question implies a proper, non trivial, and yet to be done quantisation of the only action we dispose of, that is
$\left(\int_{\rm bulk}\sqrt{-g_5}({\cal R}+\Lambda)d^5X +\kappa\int_{\rm brane}\sqrt{-g_4}{\cal L}_m d^4x\right)$.

\bigskip\medskip
\noindent
{\csc Acknowledgements}
\medskip
I gratefully thank Tomas Dolezel and Joseph Katz with whom most of the results summarized here have been first obtained as well as Cedric Deffayet and Gilles Esposito-Farese
for illuminating discussions.

\bigskip\bigskip
\noindent
{\csc References}
\bigskip
\item{[1]} L. Randall, R. Sundrum, Phys. Rev. Lett. 83 (1999) , 4690

\item{[2]} J. Garriga, T. Tanaka,  Phys. Rev. Letters 84 (2000) 2778 ;
 C. Csaki, J. Erlich, T.J. Hollowood, Y. Shirman, Nucl. Phys. B581 (2000) 309; S.B. Giddings, E. Katz, L. Randall, JHEP
0003 (2000) 023; C. Csaki, J. Erlich, T.J. Hollowood, Phys. Rev. Lett. 84 (2000) 5932; C. Csaki, J. Erlich, T.J. Hollowood,
Phys. Lett. B481 (2000) 107; I.Ya. Aref'eva, M.G. Ivanov, W. Muck, K.S. Viswanathan, I.V. Volovich, Nucl. Phys. B590 (2000)
273;  Z. Kakushadze, Phys. Lett. B497 (2000) 125

\item{[3]} P. Bin\'etruy, C. Deffayet, U. Ellwanger,  D. Langlois,  Phys. Lett. B477, 285 (2000);
 P. Kraus, JHEP 9912 (1999) 011, S. Mukoyama,  Phys. Lett. B473 (2000) 241;
 D.N. Vollick, C.Q.G. 18 (2001) 1; D. Ida, JHEP 0009 (2000) 014;  S. Mukohyama, T. Shiromizu, K. Maeda, Phys. Rev.
D62 (2000) 024028;  R. Maartens, Phys. Rev. D62 (2000) 084023; P. Kanti, K.A. Olive, M. Pospelov, Phys. Lett. B468
(1999) 31

\item{[4]} see N. Deruelle, ``Cosmological perturbations of an expanding brane in an anti-de Sitter bulk", Contribution to the Proceedings of the Porto JENAM conference, and
references therein.

\item{[5]} N. Deruelle, T. Dolezel, ``Grance versus shell cosmologies in Einstein and Einstein Gauss-Bonnet theories". gr-qc/0004021~; N. Deruelle, T. Dolezel, J. Katz,
``Perturbations of brane world", hep-th/0010215~; N. Deruelle, J. Katz, ``Gravity on branes", gr-qc/0104007~;  N. Deruelle, T. Dolezel, ``On linearised gravity in the
Randall-Sundrum scenario", gr-qc/0105118~; N. Deruelle, ``Stars on branes", gr-qc/0111065

\end